\documentclass[11pt]{article}
\usepackage{amsmath,amssymb,amsfonts,bbm}
\usepackage{multirow}
\usepackage[table,xcdraw]{xcolor}
\usepackage{graphicx}
\usepackage{caption}
\usepackage{subcaption}
\usepackage[export]{adjustbox}
\DeclareGraphicsExtensions{.pdf,.png,.jpg}
\DeclareCaptionFormat{subfig}{\figurename~#1#2#3}
\DeclareCaptionSubType*{figure}
\newcommand{\be}{\begin{equation}}
\newcommand{\ee}{\end{equation}}
\newcommand{\bea}{\begin{eqnarray}}
\newcommand{\eea}{\end{eqnarray}}
\newcommand{\ba}{\begin{array}}
\newcommand{\ea}{\end{array}}
\newcommand{\hone}{h_{1,1}}
\newcommand{\htwo}{h_{2,1}}
\newcommand{\E}{\mathcal{E}}

\newcommand{\N}{\mathcal{N}}

\newcommand{\A}{\mathcal{A}}

\long\def\symbolfootnote[#1]#2{\begingroup%
\def\thefootnote{\fnsymbol{footnote}}\footnote[#1]{#2}\endgroup}

\begin{document}

\thispagestyle{empty}\vspace{40pt}

\hfill{}

\vspace{128pt}

\begin{center}
    \textbf{\Large Geodesic structure of five-dimensional\\non-asymptotically flat 2-branes}\\
    \vspace{40pt}

    Jesse Chandler\symbolfootnote[1]{\tt jesse.chandler@cortland.edu, chanman088@gmail.com} and Moataz H. Emam\symbolfootnote[2]{\tt moataz.emam@cortland.edu}

    \vspace{12pt}   \textit{Department of Physics}\\
                    \textit{SUNY College at Cortland}\\
                    \textit{Cortland, New York 13045, USA}\\
\end{center}

\vspace{40pt}

\begin{abstract}
We study the geodesics of a five dimensional non-asymptotically flat dilatonic 2-brane. Although the metric's warp function diverges logarithmically in the far field region of the transverse space, the curvature does not. The brane has two naked singularities, the usual central singularity and a circular singularity at the radius where the warp function vanishes. This creates two causally disconnected regions of the transverse space. Using the methods of energy conservation and effective potentials we study the null and timelike orbital and radial geodesics in both regions and show that they exhibit opposing energy requirements.
\end{abstract}

\newpage

%%%%%%%%%%%%%%%%%%%%%%%%%%%%%%%%%%%%%%%%%%%%%%%%%%%%%%%%%%%%%%%%%%%%%%%%%%

%\tableofcontents

\vspace{15pt}

\pagebreak

\section{Introduction}

The concept of asymptotic flatness in physics is a time-honored principle usually imposed on gravitational solutions as a way to define isolated and closed systems. Nevertheless, non-asymptotically flat spacetime solutions exist in various theories of gravity, supergravity and string theory. There are too many of them to simply dismiss as nonphysical, and yet very little literature exists on their properties and interpretation (for example see Refs. \cite{Mazharimousavi:2010xt,Clement:2005vn, Slavov:2012mv, Clement:2004yr, Ghosh:2005aq, Ghosh:2002ut, Chatterjee:2013ova, Chan:1995fr, Yazadjiev:2005du, Kar:1996fx, Fabris:2012qe, Banerjee:2012aja}). Some of these exhibit nonvanishing curvature at infinity, while others, while endowed with non-asymptotically flat metrics, are well behaved at infinity. To our knowledge, there does not seem to be any literature on the geodesic structure of either types, which is bound to be interesting, and hence our work purports to fill an important gap in our understanding of such spacetimes. In our case study here, we investigate the geodesics of a 2-brane in five dimensional $\N=2$ supergravity theory endowed with a metric that diverges logarithmically at radial infinity \cite{Emam:2005bh}, although its curvature does not. Far from being an exotic solution designed specifically to be non-asymptotically flat, this 2-brane arises from the dimensional reduction of M2-branes over a Calabi-Yau submanifold. Furthermore, its properties are very similar to another non-asymptotically flat 2-brane, found in the same work, that arises from four intersecting M5-branes wrapped over special Lagrangian cycles of the Calabi-Yau space, which we plan to study in future work. The fact that these solutions are directly related to M-branes in $D=11$ is another reason why it is important to study their properties and geodesic and causal structures; a task we begin in this paper.

The 2-brane studied here carries a real charge that can be either positive or negative, giving two different metrics, and both have the interesting property that they change signature over two regions of the two dimensional space transverse to the brane, introducing what seems to be two extra time dimensions\footnote{This kind of behavior has been previously investigated, under different contexts, by several authors, such as Ref. \cite{Sakharov:1984ir}.}. For the positively charged solution, this is the inner region, while for the negative charge it is the outer region. In our case, the two regions are causally disconnected, and the effect their presence has on the geodesics is that they seem to be inverted with respect to each other in terms of the allowed test particle energies. While it is possible to dismiss the ``inverted'' regions as unphysical, or at least abandon the interpretation of geodesics as particle trajectories, they do arise as a direct result of the logarithmic nature of the metric, which in turn follows from the dimensional reduction of M-branes.

\section{General properties}\label{Section I}

The 2-brane subject of this study was originally found in Ref. \cite{Emam:2005bh} and is the result of the dimensional reduction of a single M2-brane down to five dimensions over a rigid Calabi-Yau manifold (Hodge numbers $\hone=\htwo = 0$). The brane's metric in the Einstein frame
\be
    ds_5^2  =  - dt^2  + dx^2  + dy^2  + f\left( r \right)\left( {dr^2  + r^2 d\theta ^2 } \right)
\ee
is characterized by the warp function $f\left( r \right)=m + q\ln \left( r \right)$, being the solution of the radial two-dimensional Laplace equation in the transverse space $\left(r,\theta\right)$. The real integration constant $m$ is the value of $f$ at $r=1$. Without loss of generality, we will set $m=1$. The constant $q$ may in principle acquire positive or negative values and we will explore both cases. It can be argued that the brane's mass density is in fact proportional to $q^2$, as briefly outlined below. The brane couples to the four scalar fields of the universal hypermultiplet, and hence it is a dilatonic solution. In fact, the warp function is directly related to the dilaton as discussed in Sec. \ref{StringFrame}.

The warp function $f\left( r \right)$ bestows on the brane two interesting properties: the first is the non-asymptotically flat nature that inspired this work in its entirety, and the second is the fact that it causes the metric signature to change from ``$-, +, +, +,+$'' to the unusual ``$-,+,+,-,-$'' at the radius $r=R_b  = e^{ - \frac{1}{q}}$. So a question that arises early on is what exactly the physical interpretation of that circular ``barrier'' $r=R_b$ is. To answer this question, we investigate the usual geometric properties: the Riemann and Ricci tensors, the Ricci scalar and the Einstein tensor respectively:
\bea
    R_{\,\,\,\theta \theta r}^r  &=&  - \frac{{q{}^2}}{{2f^2 }},\quad \quad R_{\,\,\,r\theta r}^\theta   = \frac{{q{}^2}}{{2r^2 f^2 }} \nonumber\\
    R_{rr}  &=& \frac{{q{}^2}}{{2r^2 f^2 }},\quad \quad R_{\theta \theta }  = \frac{{q{}^2}}{{2f^2 }} \nonumber\\
    R &=& \frac{{q{}^2}}{{r^2 f^3 }}\nonumber\\
    G_{tt}  &=& \frac{{q{}^2}}{{2r^2 f^3 }} =  - G_{xx}  =  - G_{yy}.
\eea

We note that all of these are in fact well behaved at radial infinity, while they diverge at $r=R_b$, where $f \rightarrow 0$. The circle $r=R_b$ is then a naked spacetime singularity. It separates the transverse space into two causally disconnected regions $r<R_b$ and $r>R_b$. The geodesics are then expected to be discontinued at that boundary. Now since $f\left( r \right)$ diverges at infinite $r$, it is rather difficult to rigorously define the mass density of the brane; however, since it is necessarily proportional to $G_{tt}$, one sees that it is dependent on $q^2$, so the brane's mass density should be positive for any real value of $q$. If one wishes to explore this in more detail, the so-called `quasilocal mass' may be investigated \cite{Brown:1992br}. The geodesic equations
\be
    \frac{{d^2 x^\sigma  }}{{d\lambda ^2 }} + \Gamma _{\mu \nu }^\sigma  \frac{{dx^\mu  }}{{d\lambda }}\frac{{dx^\nu  }}{{d\lambda }} = 0, \quad\quad \mu ,\nu ,\sigma  = 0, \ldots ,4
\ee
lead to
\bea
    \ddot r &=& \left( {q + 2f} \right)\frac{{r\dot \theta ^2 }}{{2f}} - \frac{q}{{2rf}}\dot r^2  \nonumber\\
    \ddot \theta  &=&  - \frac{{\dot r\dot \theta }}{{rf}}\left( {q + 2f} \right), \label{eom}
\eea
where an overdot represents a derivative with respect to $\lambda$: an arbitrary affine parameter in the null case, or the proper time in the timelike case. Now consider the Lagrangian $L$ of the geodesics:
\be
    2L = g_{\mu \nu } \dot x^\mu  \dot x^\nu   =  - \dot t^2  + f\dot r^2  + r^2 f\dot \theta ^2.\label{lagrangian}
\ee

Clearly, the parameter $\theta$ is cyclic, signalling the first integral
\be
    \frac{{\partial L}}{{\partial \dot \theta }} = r^2 f\dot \theta  = \ell \label{firstintegral}
\ee
where the constant $\ell$ is the angular momentum parameter. This can be used to further reduce (\ref{eom}) to the single equation
\be
    \ddot r = \left( {q + 2f} \right)\frac{{\ell^2 }}{{2r^3 f^3 }} - \frac{q}{{2rf}}\dot r^2. \label{eomII}
\ee

The nature of $r=R_b$ is immediately obvious from (\ref{eom}) or (\ref{eomII}) since $\ddot r, \ddot \theta \propto f^{-1}$ and as such diverge at $R_b$. As the equations of motion are highly nonlinear, we will attempt mostly numerical solutions with various initial conditions. For the simpler case of the radial geodesics ($\ell=0$), it is actually possible to find an analytical solution, as we will see.

\section{Orbital effective potential}\label{potential}

It is quite instructive to first study the Newtonian effective potential. We will see that it has an interesting anomalous behavior as a direct consequence to the signature flipping singularity $R_b  = e^{ - \frac{1}{q}}$. The other first integral of (\ref{lagrangian}) is found by first calculating the conserved temporal conjugate momentum
\be
    \frac{{\partial L}}{{\partial \dot t}} =  - \dot t = E
\ee
and substituting in the normalization condition
\be
    g_{\mu \nu } \dot x^\mu  \dot x^\nu   =  - \dot t^2  + f\dot r^2  + r^2 f\dot \theta ^2  = \varepsilon,
\ee
where $\varepsilon$ is zero for null geodesics and $-1$ for timelike geodesics. This gives
\be
    \E = E^2 + \varepsilon = f\dot r^2  + \frac{{\ell^2 }}{{r^2 f}},\label{energy}
\ee
where the constant $\E$ is taken to define the total energy. Now, since $\dot r^2$ cannot be allowed to become negative, we interpret the second term of the right-hand side of (\ref{energy}) as the Newtonian effective potential, so
\be
    V_{eff}  = \frac{{\ell^2 }}{{r^2 f}}.\label{Veff}
\ee

\begin{figure}[!ht]
  \begin{subfigure}[b]{.5\linewidth}
    \centering
    \includegraphics[scale=0.65]{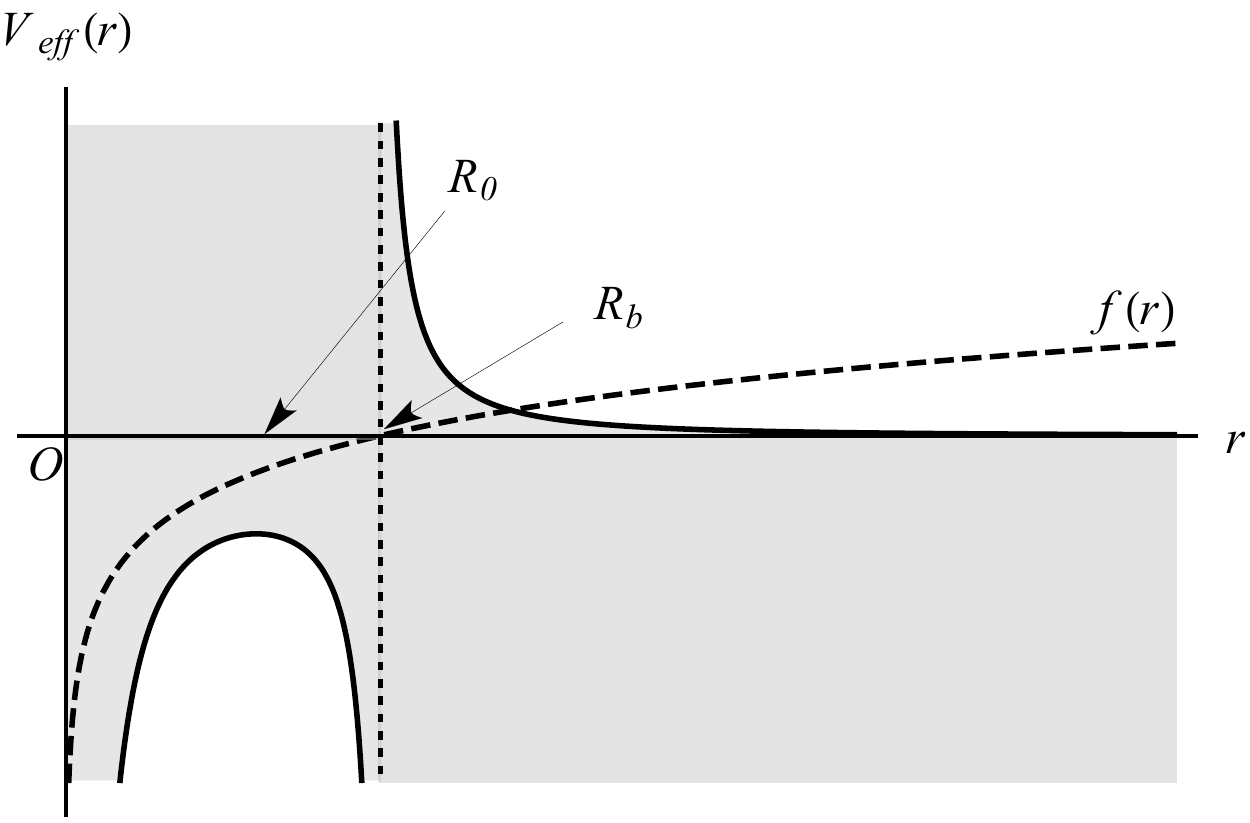}
%   \rule{4cm}{3cm}
    \caption{$q>0$}
    \label{Veffective1}
  \end{subfigure}%
  \begin{subfigure}[b]{.5\linewidth}
    \centering
    \includegraphics[scale=0.65]{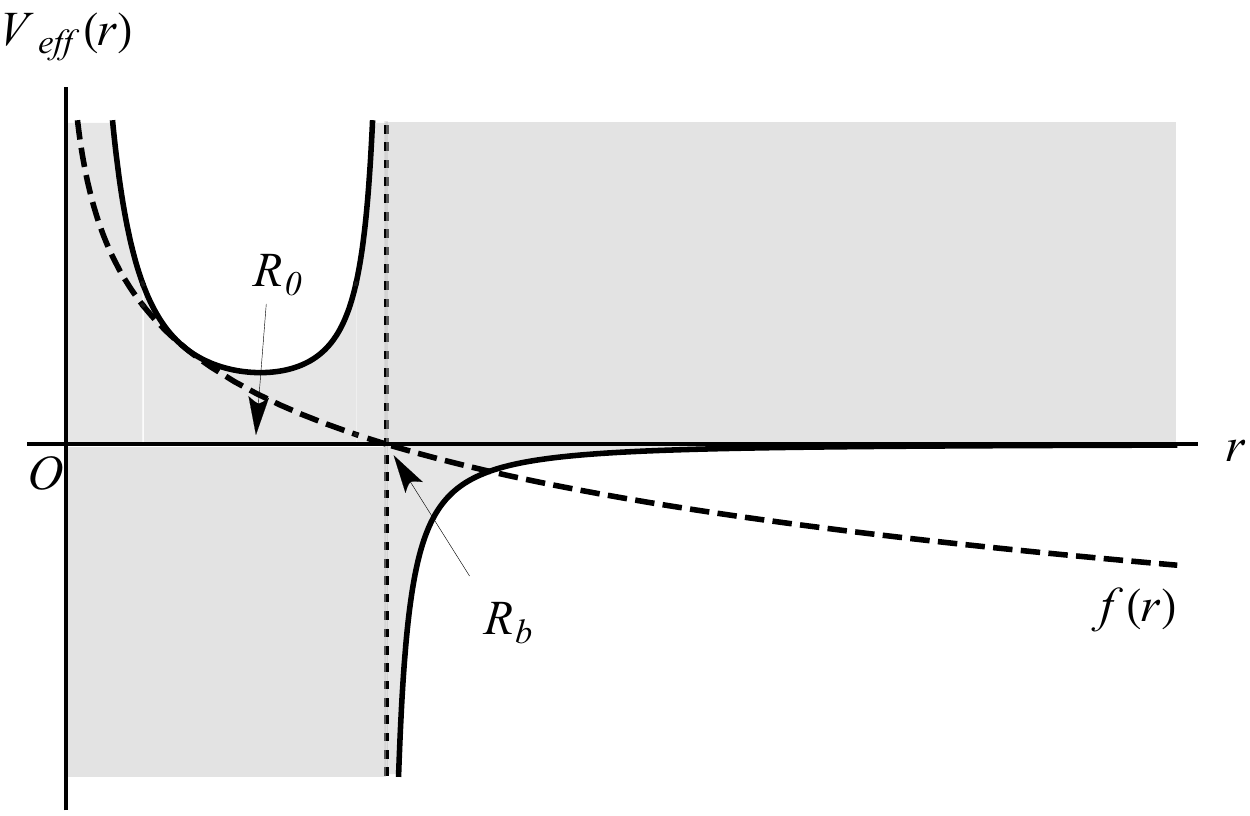}
%   \rule{4cm}{3cm}
    \subcaption{$q<0$}
    \label{Veffective2}
  \end{subfigure}
  \caption{The effective potentials and the warp function (not to scale).}
  \label{fig:1}
\end{figure}

Note, however, that the presence of $f$ allows the kinetic term $f\dot r^2$ to become negative which results in an upside-down effective potential in the inner region $r<R_b$ (this is for the case $q>0$ -- for the case $q<0$, the upside down potential occurs at $r>R_b$). Essentially, the quantity
\be
    \dot r^2  = \frac{\E-V_{eff}}{f}
\ee
must remain positive for all allowed geodesics. For the outer region $r>R_b$ ($q>0$) this is not a problem since both $f$ and $V_{eff}$ are positive so geodesics can exist for all values $\E \ge V_{eff}$. On the other hand, in the inner region $r<R_b$, both $f$ and $V_{eff}$ become negative which implies that the allowed geodesics must have values $\E \le V_{eff}$. In other words the effective potential is inverted for the inner regions, as shown in Fig. (\ref{Veffective1}), where we have also included a plot of $f\left(r\right)$ for reference. For the case $q<0$, Fig. (\ref{Veffective2}), the inner region is where the potential is ``right-side up'' and the argument is reversed. Since the shaded regions are forbidden, geodesics in the inner region can never connect to infinity and vice versa. As we will see, radial ``infall'' geodesics will exhibit similar properties, as would be expected.

\section{Stability of circular orbits}\label{circular}

Figures \ref{Veffective1} and \ref{Veffective2} clearly show the existence of circular orbits. The value of the radius $R_0$ may be found in the usual way by either setting
\be
    V'_{eff}  = \frac{{dV_{eff} }}{{dr}} =  - \frac{{\ell^2 }}{{r^3 f^2}}\left( {q + 2f} \right)
\ee
to zero and solving for $r=R_0$, or by setting $\ddot r=\dot r=0$ in (\ref{eomII}). Both of these give
\bea
    q + 2f\left( R_0 \right) &=& 0 \nonumber\\
    \rightarrow\quad R_0 &=& e^{ - \left( {\frac{1}{2} + \frac{1}{q}} \right)}.
\eea

Usually, there are two methods of exploring the stability of circular orbits. The first is the standard second derivative test
\be
    \left. {V''_{eff} } \right|_{r = R_0}  = \frac{{l^2 }}{{r^4 f^2 }}\left( {5q + 6f + 2\frac{{q^2 }}{f} } \right)_{r = R_0} \left\{ {\begin{array}{*{20}c}
   { >0\quad \quad \textrm{stable}}  \\
   { =0\quad \quad \textrm{critical}}  \\
   { <0\quad \quad \textrm{unstable}}  \\
    \end{array}} \right.
\ee

In this case, however, this method leads to the incorrect result for $q>0$, since it does not test the actual dynamical stability of the orbit but rather simply tests whether the potential has a minimum or a maximum. The true stability is tested by perturbatively disturbing the circular orbit and exploring whether it leads to an oscillation about $R_0$ or not. This is done by plugging $r=R_0+\epsilon$, where $\epsilon<<R_0$, in (\ref{eomII}), expanding and ignoring $O\left( {\epsilon ^2 } \right)$ terms and higher, giving
\be
    \ddot \epsilon  + \left( {\frac{{8\ell^2 }}{{q^2 R_0^4 }}} \right)\epsilon  = 0.\label{oscilator}
\ee

This is of course the standard simple harmonic differential equation with angular frequency
\be
    \omega  = \left| {\frac{{2\sqrt 2 \ell}}{{qR_0^2 }}} \right|.\label{omega}
\ee

The fact that $\omega ^2$ is positive for both $q=\pm 1$ indicates that the circular orbit is stable under small perturbations even for the case $q>0$ where the potential is clearly upside down.

\section{Orbital geodesics}\label{orbital}

\begin{figure}[hp]
  \begin{subfigure}[b]{.5\linewidth}
    \centering
    \includegraphics[scale=0.5]{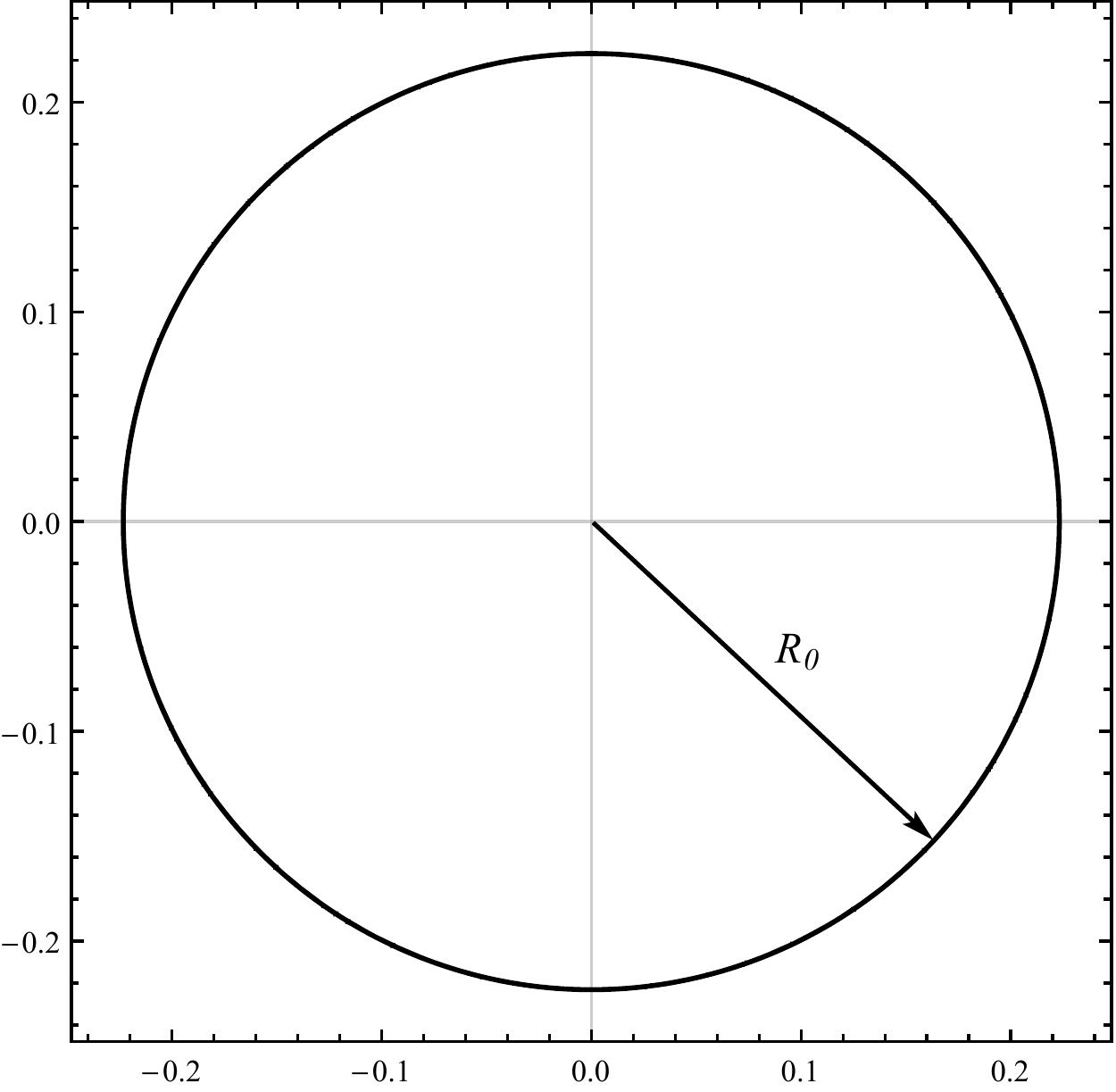}
%   \rule{4cm}{3cm}
    \caption{The circular orbit}
    \label{1}
  \end{subfigure}
  \begin{subfigure}[b]{.5\linewidth}
    \centering
    \includegraphics[scale=0.5]{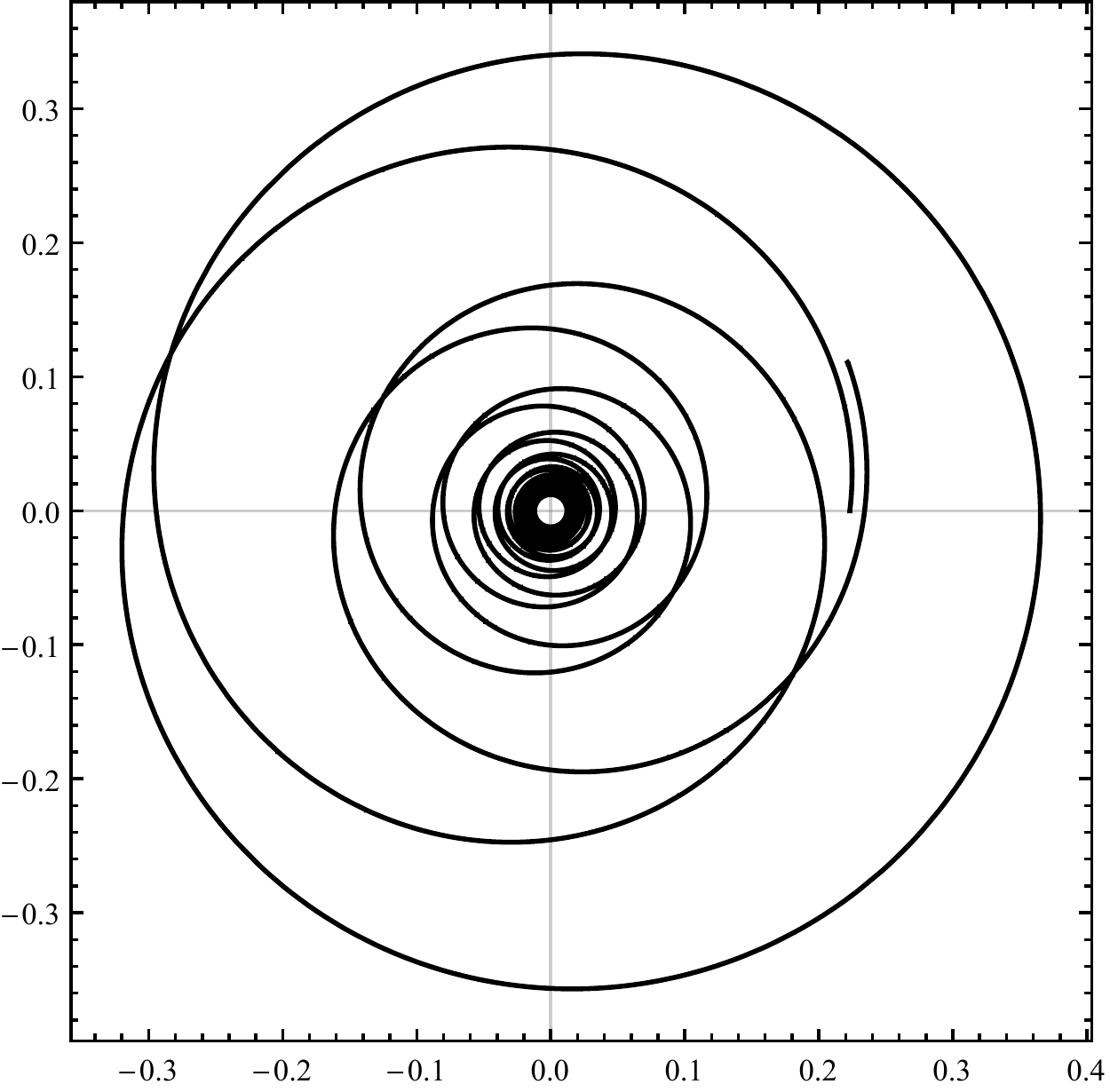}
%   \rule{4cm}{3cm}
    \subcaption{An inner orbit with low energy}
    \label{2}
  \end{subfigure}
  \begin{subfigure}[b]{.5\linewidth}
    \centering
    \includegraphics[scale=0.5]{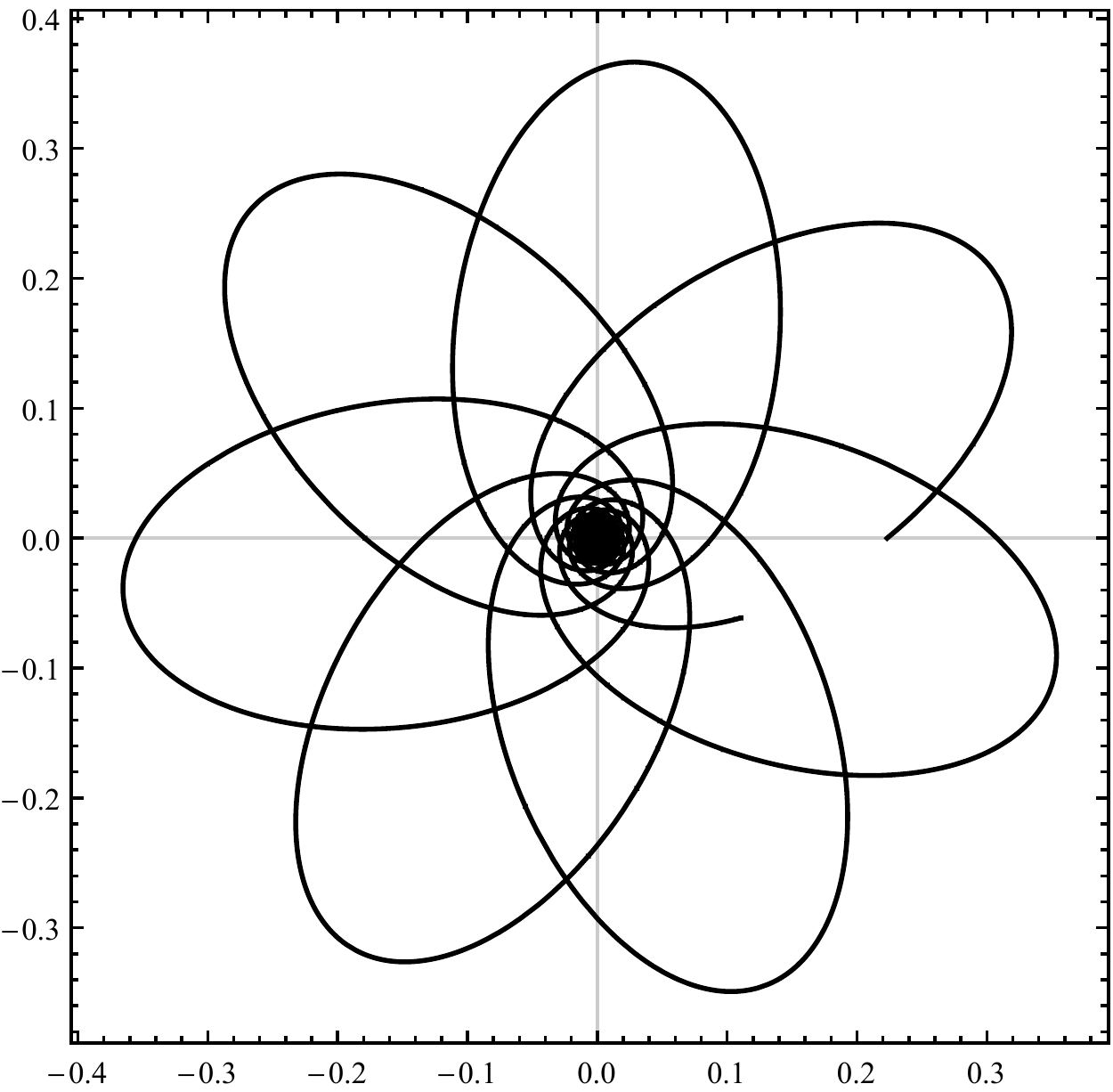}
%   \rule{4cm}{3cm}
    \caption{An inner orbit with high energy}
    \label{3}
  \end{subfigure}
  \begin{subfigure}[b]{.5\linewidth}
    \centering
    \includegraphics[scale=0.4]{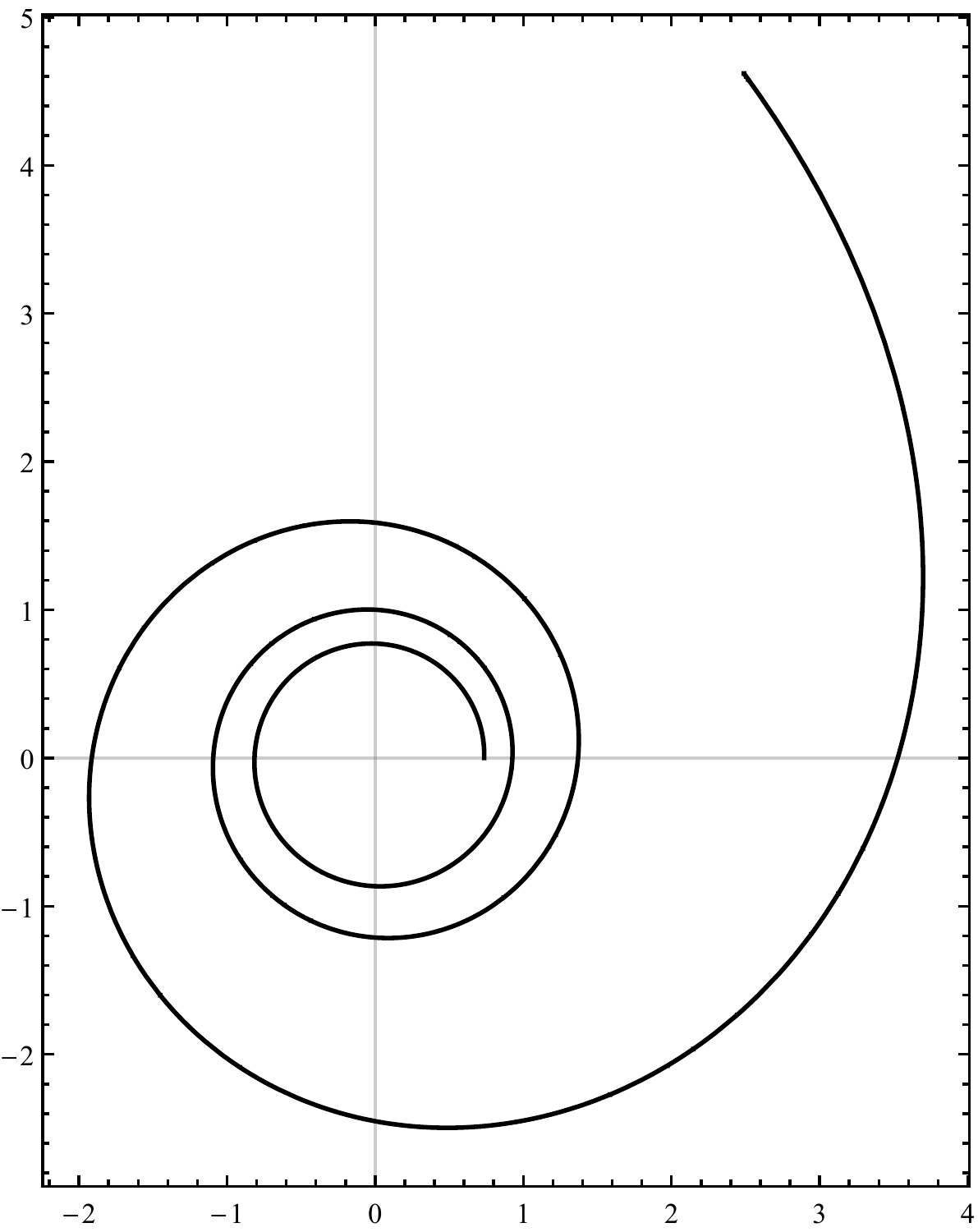}
%   \rule{4cm}{3cm}
    \caption{An outer region geodesic: $\dot r_i > 0$ }
    \label{4}
  \end{subfigure}
  \begin{subfigure}[b]{.5\linewidth}
    \centering
    \includegraphics[scale=0.5]{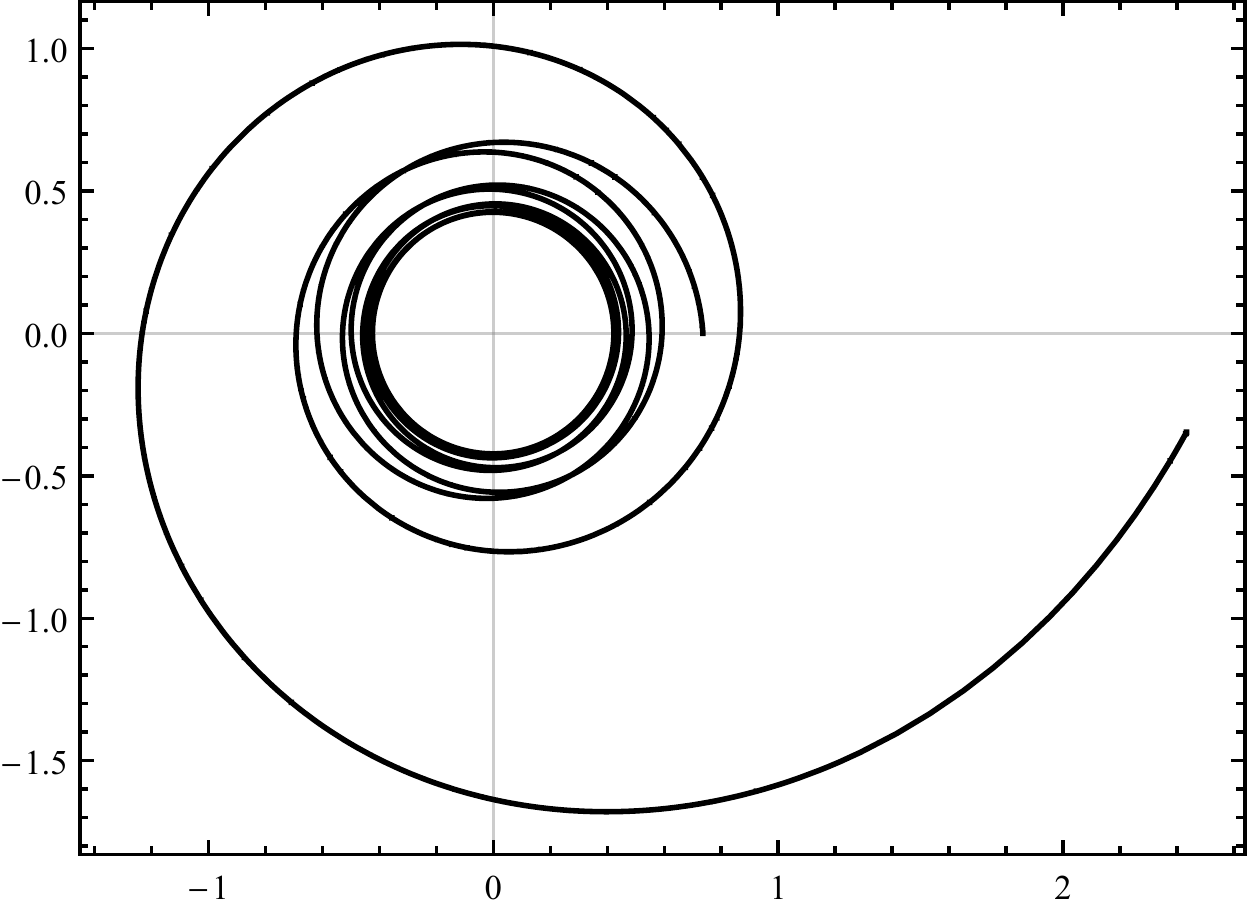}
%   \rule{4cm}{3cm}
    \caption{An outer region geodesic: $\dot r_i < 0$}
    \label{5}
  \end{subfigure}
  \begin{subfigure}[b]{.5\linewidth}
    \centering
    \includegraphics[scale=0.4]{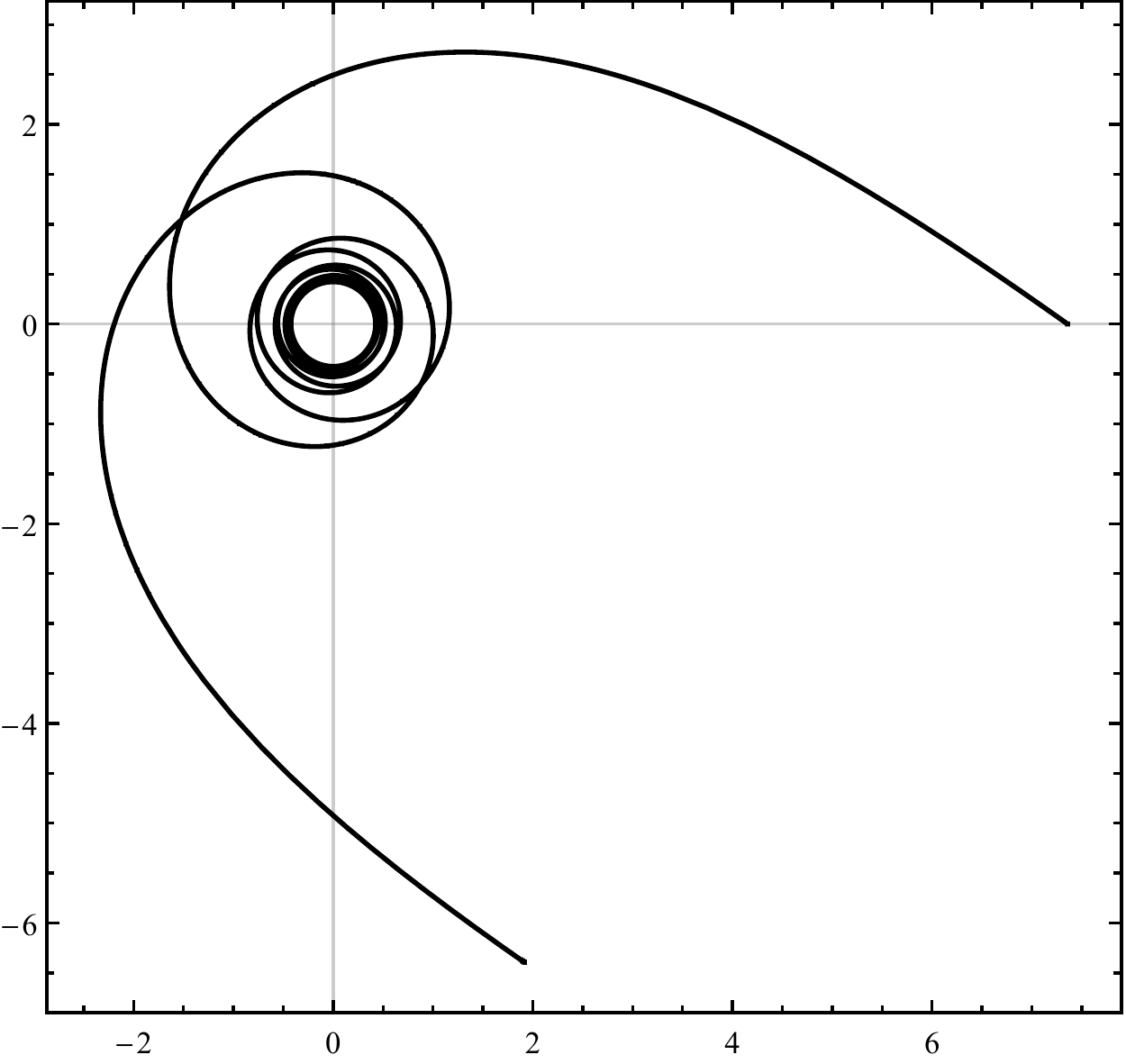}
%   \rule{4cm}{3cm}
    \caption{A slingshot geodesic: $\dot r_i <0$, $r_i \rightarrow \infty $}
    \label{6}
  \end{subfigure}
  \caption{Results of numerically solving equations (\ref{eom}) for $q=+1$. The $q=-1$ plots are identical, except that the scale is adjusted by a factor of $e^2$. For example, the radius of the circular orbit for $q=+1$ is $R_0=0.2231$, while for the $q=-1$ case it is $R_0=0.2231 \times e^2=1.6487$ and so on.}
  \label{fig:2}
\end{figure}

\begin{figure}[!ht]
  \begin{subfigure}[b]{.5\linewidth}
    \centering
    \includegraphics[scale=0.4]{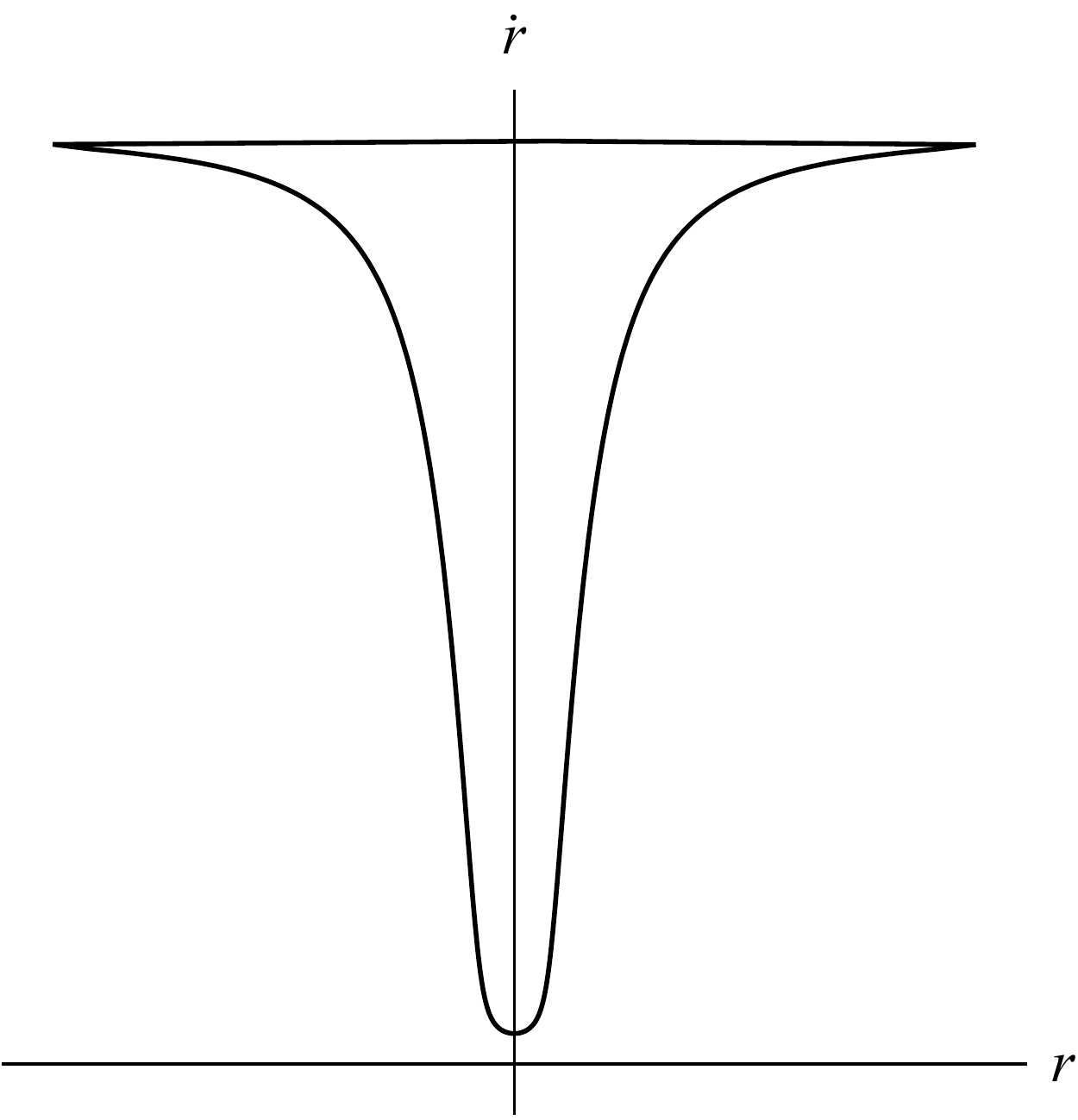}
%   \rule{4cm}{3cm}
    \caption{The phase diagram for figure (\ref{2})}
    \label{P1}
  \end{subfigure}%
  \begin{subfigure}[b]{.5\linewidth}
    \centering
    \includegraphics[scale=0.4]{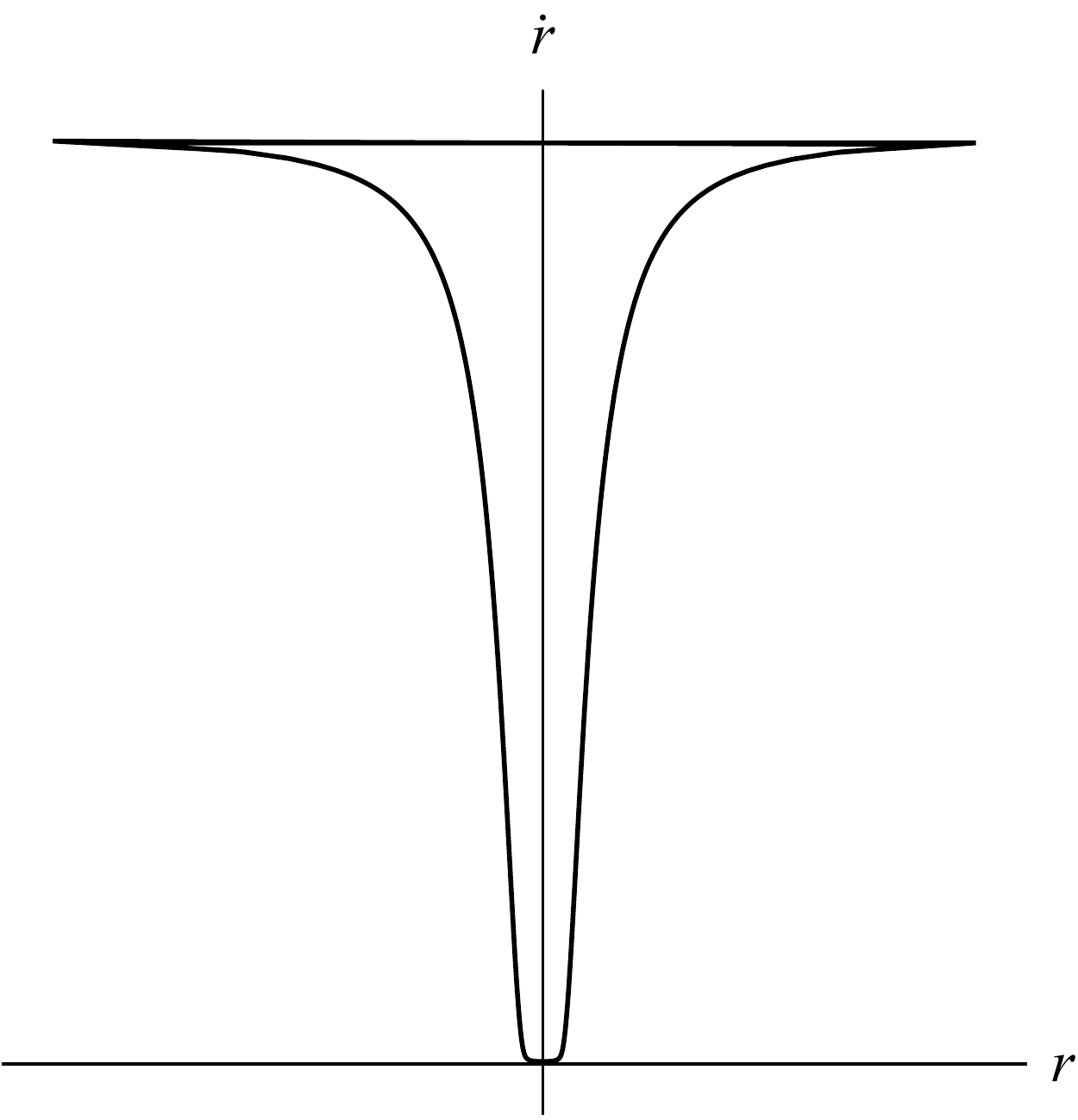}
%   \rule{4cm}{3cm}
    \subcaption{The phase diagram for figure (\ref{3})}
    \label{P2}
  \end{subfigure}
  \caption{Phase diagrams $\dot r$ vs $r$ for the inner orbits in Figs. \ref{2} and \ref{3}. These are for the case $q=+1$. The ones for the case $q=-1$ are identical up to a scale as well.}
  \label{fig:3}
\end{figure}

We now attempt numerical solutions to Eqs. (\ref{eom}) classified by different values of the total energy $\E$. Another peculiar consequence of the warp factor $f$ is the possibility of getting negative values (left handed) for the initial angular speed $\dot\theta_i$ corresponding to positive angular momenta $\ell$ (right handed), via (\ref{firstintegral}). We then set $\ell=\pm 1$ as appropriate to always give positive values for $\dot\theta_i$. The energies $\E$ are chosen as multiples of the energy of circular orbits
\be
    \E_0= \frac{{ - 2\ell^2 }}{{qR_0^2 }}
\ee
and the initial radial velocity is calculated using
\be
    \dot r_i  =  \pm \sqrt {\frac{{\E  - V_{eff} \left( {r_i ,\ell } \right)}}{{f\left( {r_i } \right)}}}
\ee
where $r_i$ is the initial radius. All plots start at the horizontal axis, \emph{i.e.} the initial angle is $\theta_i=0$. The results are collectively shown in Table \ref{Table1}. Figures \ref{fig:2} represent the null/timelike geodesics for $q=+1$. The plots for the case $q=-1$ are very similar in form, just larger by a factor of $e^{{2 \mathord{\left/ {\vphantom {2 {\left| q \right|}}} \right. \kern-\nulldelimiterspace} {\left| q \right|}}} $. Finally, the inner regions' geodesics can be shown to be closed by plotting their phase diagram $\dot r$ vs $r$. These are given in Fig. \ref{fig:3}.

\vspace{12 pt}

% \usepackage{multirow}
\begin{table}[hp]
\centering
\begin{tabular}{|c|c|c|c|c|c|}
\hline
$q=\pm 1$                       & Figure number & $\ell$            & $r_i$                   & $\E$                & $\dot r_i $    \\ \hline
\multirow{3}{*}{Inside region}  & \ref{1}     & \multirow{3}{*}{$\mp 1$} & \multirow{3}{*}{$R_0$}  & $\E_0$              & 0                      \\ \cline{2-2} \cline{5-6}
                                & \ref{2}     &                     &                         & $50\E_0$            & \textgreater0         \\ \cline{2-2} \cline{5-6}
                                & \ref{3}     &                     &                         & $5000\E_0$          & \textgreater0          \\ \hline
\multirow{3}{*}{Outside region} & \ref{4}     & \multirow{3}{*}{$\pm 1$} & \multirow{2}{*}{$2R_b$} & ${-\E_0 \mathord{\left/ {\vphantom {-\E_0 4}} \right. \kern-\nulldelimiterspace} 4}$                                                                                      & \textgreater0            \\ \cline{2-2} \cline{5-6}
                                & \ref{5}     &                     &                         & $-\E_0$             & \textless0              \\ \cline{2-2} \cline{4-6}
                                & \ref{6}     &                     & $20R_b\approx \infty $  & $-\E_0$             & \textless0                \\ \hline
%\multicolumn{6}{|c|}
%{\begin{tabular}[c]{@{}c@{}}
%\\ $\quad$ \includegraphics[scale=0.5]{Solutions1} $\quad\quad$
%\includegraphics[scale=0.5]{Solutions2}$\quad$\\
%\\
%\end{tabular}}                                                                                                                                                                                  %\\ \hline
\end{tabular}
\caption{The $q=\pm 1$ solutions classified by energy and initial conditions}
\label{Table1}
\end{table}

%0.2231301601 and 1.648721271

\pagebreak

\section{Radial geodesics}\label{Radial}

We now study the radial geodesics. The energy conservation relation (\ref{energy}) reduces to
\be
    \dot r^2  = \frac{\E}{f},\label{energy2}
\ee
implying that radial geodesics with positive total energy $\E$ cannot exist inside the barrier $r<R_b$ (for the case $q>0$ and the reverse argument for $q<0$). Equation (\ref{energy2}) leads to
\be
    \lambda\left( r \right) = \int {dr\sqrt {\frac{f}{\E}} }
\ee
the solution of which is
\be
    \lambda\left( r \right) = \frac{1}{2}\sqrt {\frac{f}{\E}} \left[ {2r + \frac{{\sqrt \pi R_b }}{\chi } \textmd{erfc}\left( \chi \right)} \right],\label{RadialSolution}
\ee
where
\be
    \textmd{erfc}\left( \chi \right) = \frac{2}{{\sqrt \pi  }}\int\limits_\chi^\infty  {e^{ - u^2 } du}.
\ee
is the Gauss complimentary error function and
\be
    \chi \left( r \right) = \sqrt { - \frac{f}{q}}.
\ee

For positive values of $q$, the function $\chi$ is real only for negative $f$, which is true in the region $r<R_b$. The square root ahead of (\ref{RadialSolution}) also requires $\E<0$ as in the orbital case. On the other hand, in the region $r>R_b$, positive energies are required and the overall $\lambda\left( r \right)$ is still positive because the product of the now complex $\chi$ with its own complimentary error function is real. The argument is reversed for the case $q<0$.

\begin{figure}[hp]
  \begin{subfigure}[b]{.5\linewidth}
    \centering
    \includegraphics[scale=0.45,left]{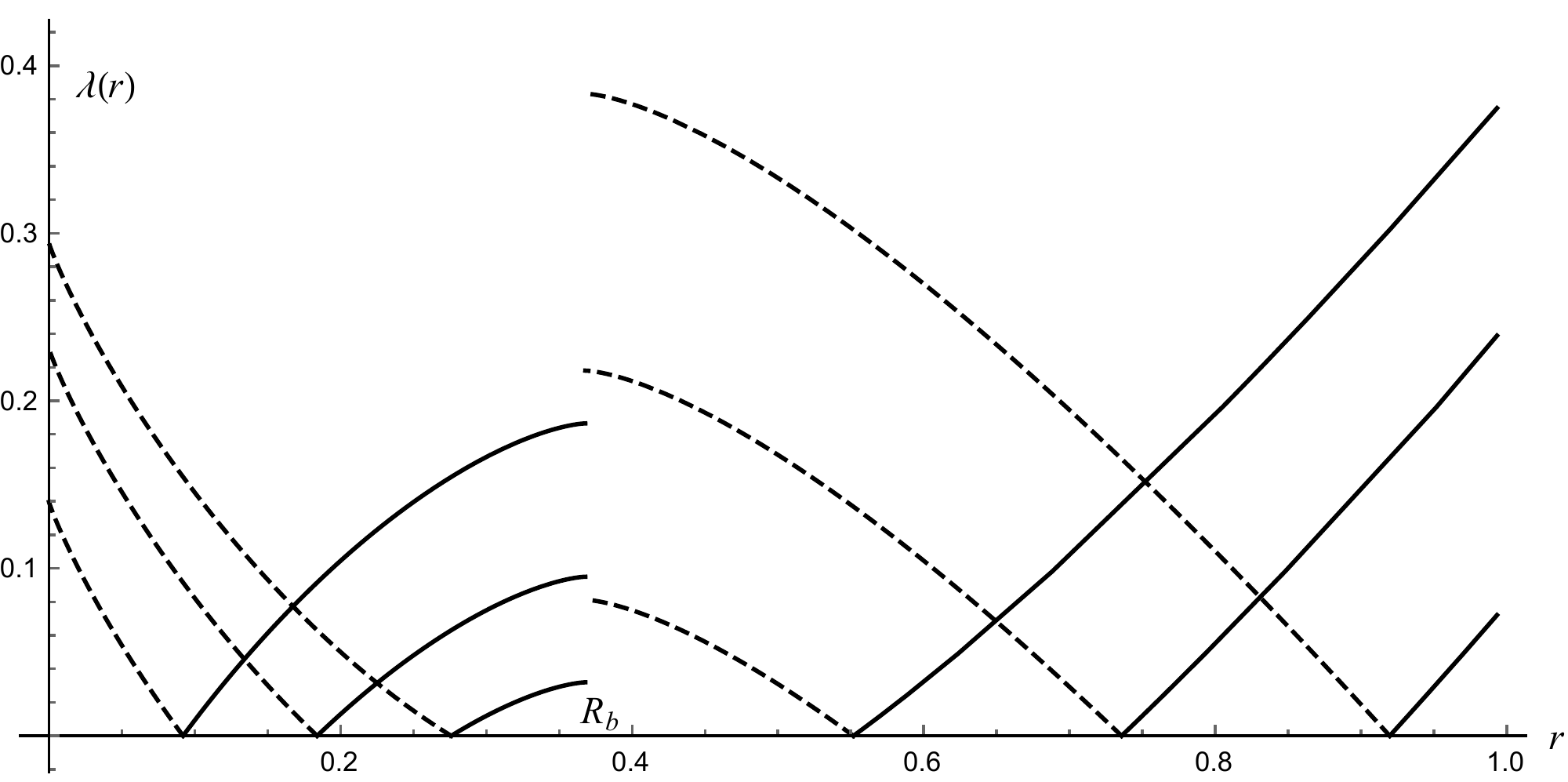}
%   \rule{4cm}{3cm}
    \caption{Radial world lines.}
    \label{fig:4a}
  \end{subfigure}
  \begin{subfigure}[b]{.5\linewidth}
    \centering
    \includegraphics[scale=0.4,right]{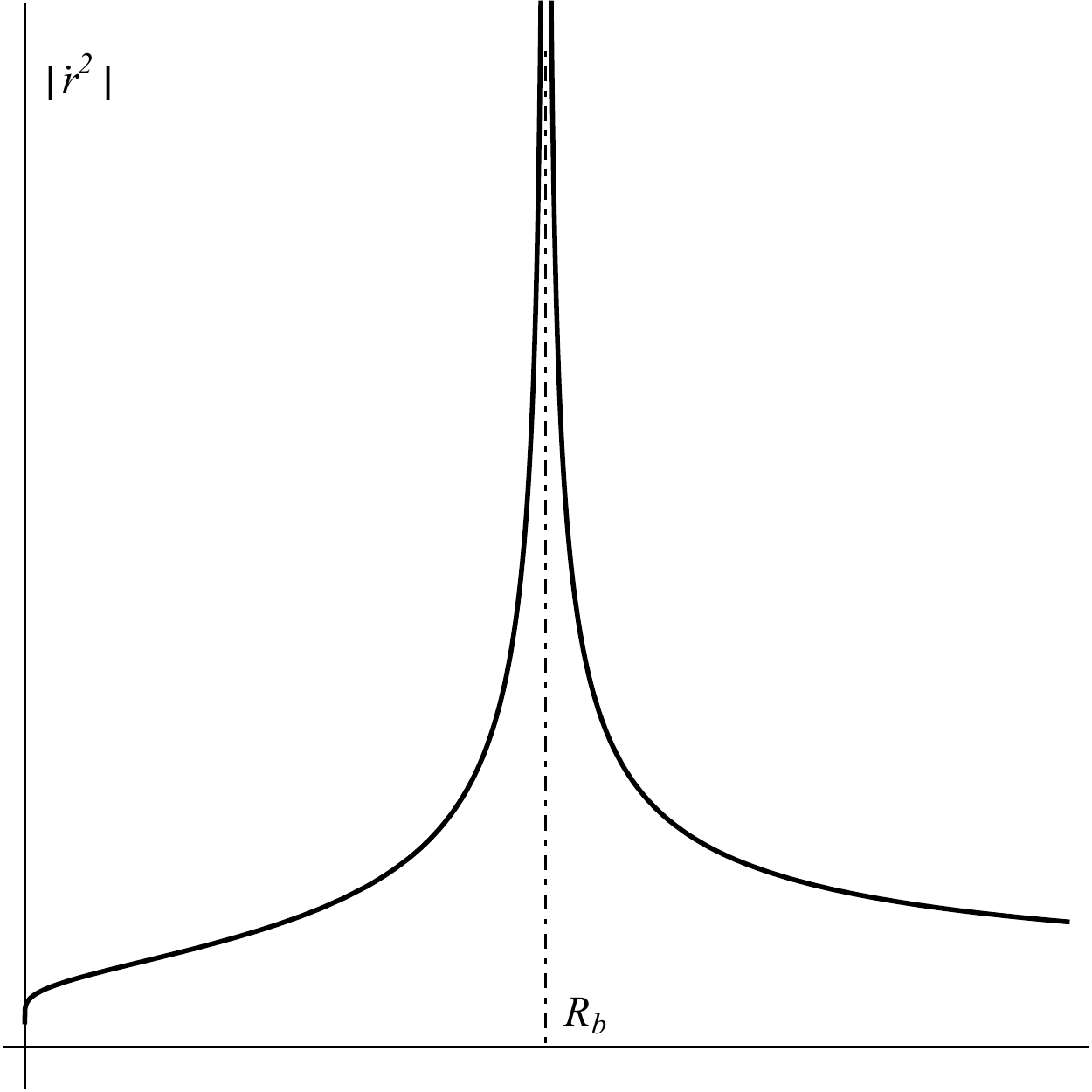}
%   \rule{4cm}{3cm}
    \subcaption{$\left| {\dot r^2 } \right|$ vs $r$}
    \label{fig:4b}
  \end{subfigure}
  \caption{Radial geodesics}
\end{figure}

Figure \ref{fig:4a} shows $\lambda\left(r\right)$ plots of six double geodesics\footnote{Light cones in the null case.} for $q=+1$, three in the inside region $r<R_b$ and three in the outside region $r>R_b$. The inner geodesics have $\E=-1$, with initial radii ${{R_b } \mathord{\left/ {\vphantom {{R_b } 4}} \right. \kern-\nulldelimiterspace} 4}$, ${{R_b } \mathord{\left/ {\vphantom {{R_b } 2}} \right. \kern-\nulldelimiterspace} 2}$, and ${{3R_b } \mathord{\left/ {\vphantom {{3R_b } 4}} \right.
 \kern-\nulldelimiterspace} 4}$. The outer geodesics have $\E=+1$ with initial radii ${{3R_b } \mathord{\left/ {\vphantom {{3R_b } 2}} \right.
 \kern-\nulldelimiterspace} 2}$, $2R_b$ and ${{5R_b } \mathord{\left/ {\vphantom {{5R_b } 2}} \right. \kern-\nulldelimiterspace} 2}$. For each point we show outward (the solid lines: $\dot r_i>0$) and inward (the dashed lines: $\dot r_i<0$) geodesics. The singularity $r=R_b$ appears naturally, causally disconnecting the inner and outer regions.

In the outer region the geodesics exhibit the expected $\dot r$ behavior: outward curves seem to ``slow down'' as they move away from the brane, while inward curves speed up as they move toward the ring singularity. In the inside region, the curves speed up toward the ring singularity but slow down toward the central singularity $r=0$. To clarify this, consider (\ref{energy2}), this function is ``symmetrical'' about $r=R_b$, in the sense that $\left| {\dot r} \right|$ gets larger as it approaches $r=R_b$ and smaller as it moves away from it, \emph{in any direction}, even toward the central singularity. This is yet another unusual property of this 2-brane. Figure \ref{fig:4b} shows a $q=+1$ plot of $\left| {\dot r^2 } \right|$ that further clarifies this point.

\vspace{12 pt}

\pagebreak

\section{The universal hypermultiplet and the string frame}\label{StringFrame}

As shown in Ref. \cite{Emam:2005bh}, the 2-brane couples to two of the four scalar fields of the universal hypermultiplet, these being the dilaton $\sigma$ and the 3-form gauge potential $\A_{012}$, while the two axions vanish. The fields are
\bea
    \sigma \left( r \right) &=&  - \ln f\left( r \right)\nonumber\\
    A_{012}&=& \pm f^{-1}.
\eea

\begin{figure}[!ht]
  \begin{subfigure}[b]{.5\linewidth}
    \centering
    \includegraphics[scale=0.6]{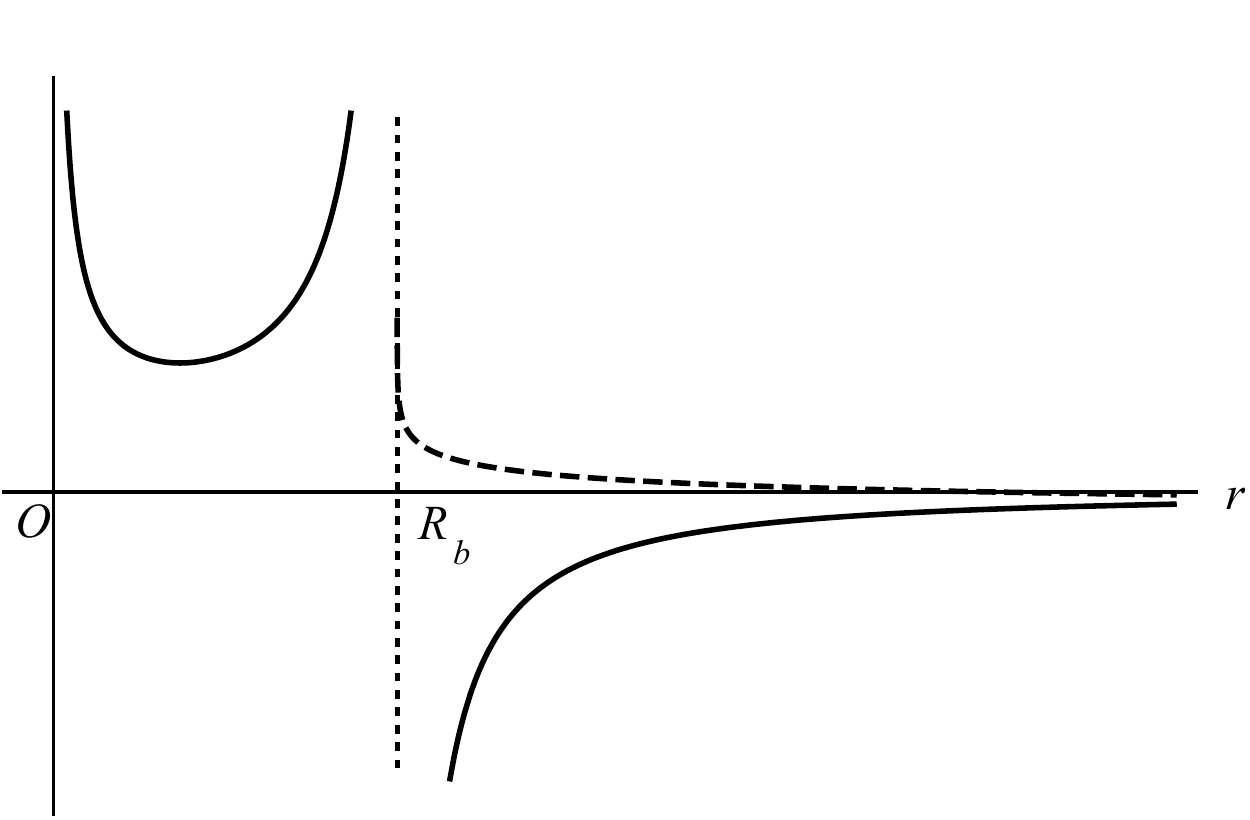}
%   \rule{4cm}{3cm}
    \label{Dilatonpos}
  \end{subfigure}%
  \begin{subfigure}[b]{.5\linewidth}
    \centering
    \includegraphics[scale=0.6]{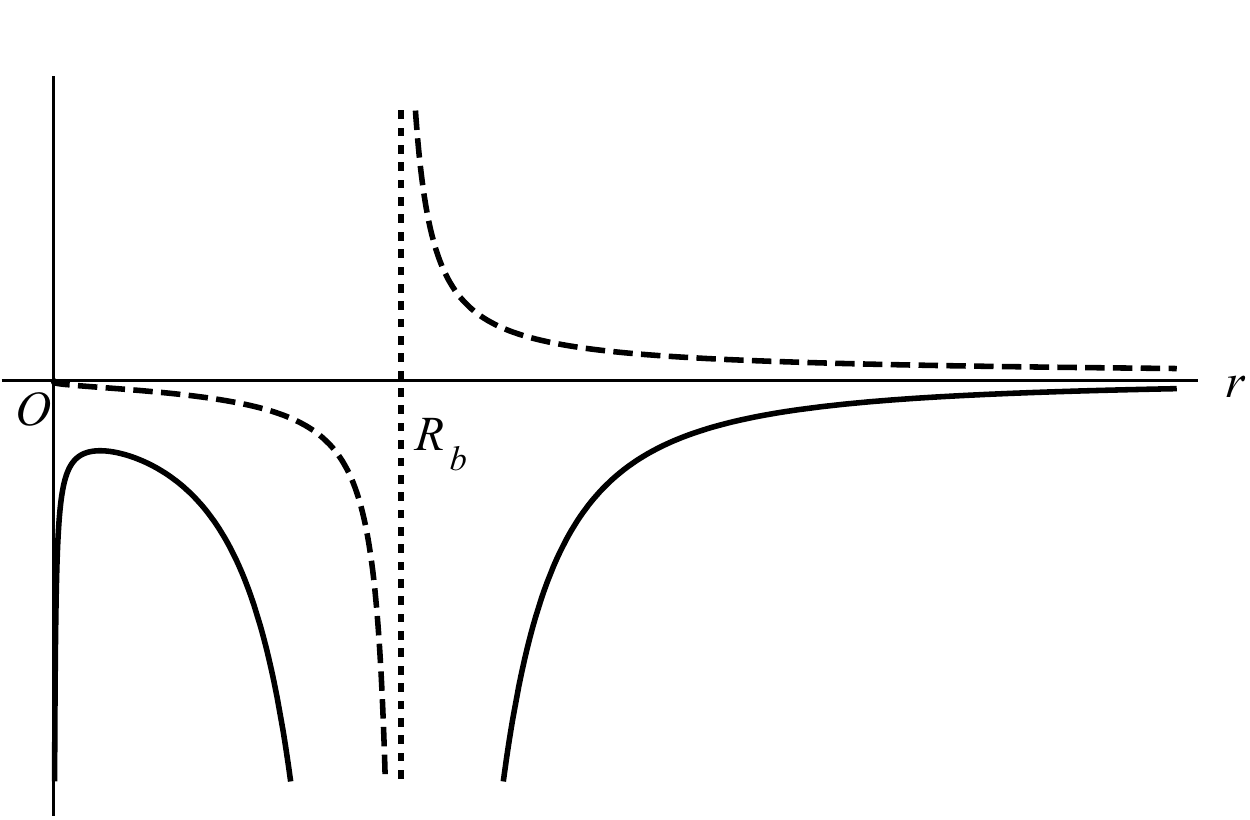}
%   \rule{4cm}{3cm}
    \label{FandApos}
  \end{subfigure}
  \caption{The hypermultiplet fields for the case $q>0$. The dashed lines represent the potentials $\sigma$ (left) and $A_{012}$ (right), while the solid lines represent the fields $d\sigma$ and $F_{r012}$ respectively.}
  \label{fig:5}
\end{figure}

Clearly, the dilaton $\sigma$ only has real values in the regions ($q>0$, $r>R_b$) and ($q<0$, $r<R_b$), \emph{i.e.} regions where the metric is positive. However, the dilaton's field strength $d\sigma$ exists everywhere and is singular at both $r=0$ and $r=R_b$, and the same is true for the 4-form field strength $F_{r012}=dA_{012}$, as shown in Fig. \ref{fig:5}. Explicitly
\bea
    d\sigma  &=&  - \frac{q}{{rf}} dr\nonumber\\
    F_{r012}  &=&  \mp \frac{q}{{rf^2 }} dr.
\eea

For completeness, we reanalyzed the results in the previous sections in the string frame, where the metric becomes
\bea
    ds_{{\rm String}\,\,{\rm frame}}^2  &=& e^{\frac{4}{3}\sigma } ds_{{\rm Einstein}\,\,{\rm frame}}^2\nonumber\\
      &=& f^{ - \frac{4}{3}} \left( { - dt^2  + dx^2  + dy^2 } \right) + f^{ - \frac{1}{3}} \left( {dr^2  + r^2 d\theta ^2 } \right).
\eea

The geodesic equations in this frame are
\bea
    \ddot t &=& \frac{{4q}}{{3rf}}\dot t\dot r,\quad\quad \ddot \theta  = \frac{{\dot r\dot \theta }}{{3rf}}\left( {q - 6f} \right)\nonumber\\
    \ddot r &=& \frac{q}{{6rf}}\left( {\dot r^2  - r^2 \dot \theta ^2 } \right) + r\dot \theta ^2  + \frac{{2q}}{{3rf^2 }}\dot t^2,\\
\eea
along with the first integrals
\be
    \dot t =  - f^{\frac{4}{3}} E, \quad\quad
    \dot \theta  = \frac{{f^{\frac{1}{3}} }}{{r^2 }}\ell.
\ee

For null geodesics, energy conservation leads to
\be
    \E = E^2  = f^{ - \frac{5}{3}} \dot r^2  + V_{eff},
\ee
where the effective potential is still defined by (\ref{Veff}). Consequently the geodesics in the string frame are similar to those in the Einstein frame and are still described, up to scaling factors, by the previous plots [Figs \ref{fig:2}, \ref{fig:3} and \ref{fig:4a}] as can be easily checked.

\section*{Conclusion}

In this work, we have studied the geodesic structure of a non-asymptotically flat 2-brane in $D=5$ supergravity. This solution is one of two branes that arise as the dimensional reduction of M-branes over a rigid Calabi-Yau manifold, the second of which we plan to explore in future work, although the fact that they have similar properties implies that they probably share similar geodesic structures (the second brane differs only in that it does \emph{not} change signature). The brane's metric contains a logarithmic warp function $f\left(r\right)$ that diverges at radial infinity. Although the curvature of the transverse space vanishes at infinity, it diverges on the brane ($r=0$) as well as at the radial distance $r=R_b$, the root of the logarithmic function. Hence there are two causally disconnected regions separated by a circular ring singularity. The warp function $f\left(r\right)$ flips sign over the boundary of these two regions introducing an interesting behavior to the geodesics' energy conservation, in that the Newtonian effective potential is upside down in one of the regions as compared to the other. Only negative energy geodesics can exist in one and positive energy geodesics in the other. The fact that the metric changes signature over this boundary raises all kinds of interesting questions as to the causal structure in the regions where there is more than one time dimension. This begs for further investigation in the future. It is also possible to argue, however, that such a signature change simply introduces a forbidden spatial region. So regions of the transverse space where the effective potential is inverted are nonphysical and nothing can exist there. This is further amplified by noting that the dilaton potential can only exist in regions where the brane carries the  standard signature. On the other hand the fields do exist everywhere, and reanalysis in the string frame showed no difference from that in the Einstein frame. Again, due to the uniqueness of such non-asymptotically flat solutions and their relation to the dimensional reduction of M-branes, this begs for further investigation before any such conclusions can be firmly established. Furthermore, although the literature contains several solutions with non-asymptotically flat metrics (with or without vanishing curvature at infinity), there does not seem to be any work done on the geodesic structure of such solutions. This paper then fills an important gap. In the future, we also plan to investigate the geodesic structure of spacetime solutions of which the curvature does not vanish at an infinite distance from the source.

\end{document}